\RequirePackage{fix-cm}
\documentclass[twocolumn,epjc3]{svjour3}  
\smartqed  
\RequirePackage{graphicx}
\usepackage{caption}
\usepackage{makecell}
\usepackage[labelformat=simple, skip=0pt]{subcaption}

\usepackage{amsmath,amssymb}

\journalname{Eur. Phys. J. C}

\begin{document}

\title{The ``$B\to K\pi$ puzzle'': A New Perspective}

\author{Adam Szabelski\thanksref{e1,addr1}
}

\thankstext{e1}{e-mail: adam.szabelski@ncbj.gov.pl}

\institute{National Centre for Nuclear Research \label{addr1}}

\date{Received: date / Accepted: date}

\maketitle

\begin{abstract}
A phenomenological analysis based on the published branching fractions and $CP$ asymmetry observables of the $B\to K\pi$, $B\to\pi\pi$ and $B\to KK$ dataset is performed. The amplitude decomposition by the  topological diagrams and the $SU(3)$ irreducible representation is used. The result of the global fit can be interpreted as a solution to the ``$B\to K\pi$ puzzle''. The obtained precision enables to test a $SU(3)$-based sum rule showing  $6\sigma$ discrepancy with the expectation value in the Standard Model scenario. To improve the obtained results the main effort should concentrate on the $B^0\to K^0\pi^0$ measurements.
\end{abstract}

\section{Introduction}
The study of the B meson decays provides the opportunity to experimentally investigate the flavour- and $CP$-violating features of the Standard Model ($SM$) \cite{Buras:2013ooa}, described by the elements of the Cabibbo-Kobayashi-\newline-Maskawa(CKM) matrix \cite{Cabibbo:1963yz,Kobayashi:1973fv}.
These explorations are performed in $B$-factories. The first results were brought by the BaBar and Belle experiments as well as the Tevatron. Today, the stage is taken over by LHCb at the Large Hadron Collider and Belle II at the KEK facility.

In the spotlight of the B meson decays the family of the $B\to K\pi$ transitions plays a key role. 
Its importance was strengthened when the ``$B\to K\pi$ puzzle'' was formulated~\cite{Buras:2003yc} pointing out that significant deviation from zero of for experimentally determined quantity 
\begin{equation}
    \Delta A_{CP} = A_{CP}(B^+\to K^+\pi^0)-A_{CP}(B^0\to K^+\pi^-)
    \label{eq:DA_CP}
\end{equation}    
cannot be explained within the $SM$ framework and that the explanation should lead to the inclusion of the new physics ($NP$) phenomena to the model.
The $B\to K\pi$ decays are dominated by the QCD penguin topologies with the tree contribution suppressed by the $|V_{ub}|$ CKM matrix element. The electro-weak ($EW$) penguin amplitudes play an important role both for $CP$ violation and the branching fraction (BF) evaluation for these reactions \cite{Buras:2003yc,Buras:2003dj,Buras:2004ub,Baek:2004rp}. The potential effects of the $NP$ are expected to manifest in the $EW$ penguin contribution.  
This effect may be observed either as a nonzero phase difference - $\phi$ - between the $EW$ penguin and the tree amplitudes, or as a deviation of their magnitude ratio - $q$ - from the value predicted by the $SM$ \cite{Neubert:1997wb}.

In this study, the approach proposed in~\cite{Baek:2004rp,Baek:2007yy,Baek:2009pa,Fleischer:2017vrb} is used. The decay amplitudes are decomposed into the Feynman graphs. Each graphical description represents a topology and includes all multi-loop diagrams that can be reduced down to the given form. To each topology, a single complex parameter is assigned. The number of the free parameters is decreased by imposing the isospin relation and by applying the $SU(3)$ decomposition~\cite{Gronau:1998fn} (for details see Section~\ref{subsect:par1}).
The parameters are fitted to the experimental results: direct and time-dependent $CP$ asymmetries and $BFs$ of the $B\to K\pi$, $B\to\pi\pi$ and $B\to KK$ decays. The contribution of the $NP$ is expected to be noticeable in either the magnitude or the phase of the $EW$ penguin amplitude. 

The same $SU(3)$ decomposition leads to a sum rule composed of rate asymmetries. It was proposed in~\cite{Gronau:2005kz} and is constructed in such a way 
 that it is small in the sense of the asymmetry range. The rate asymmetries in the sum rule differ by $d\leftrightarrow u$ and naively the sum rule should be zero.
 
The analysis presented here incorporates 
the latest results from LHCb on $B^+\to K^+\pi^0$ \cite{LHCb:2020dpr} and B decays to charged pions and kaons \cite{LHCb:2020byh} as well as recent Belle~II measurements of $B\to K\pi$~\cite{Belle-II:2023ksq}.
The results are represented as the profile plot in the $q-\phi$ plane based on the minimal $\chi^2$ of the fits. The plot demonstrates how well the $SM$ expectation value for $qe^{i\phi}$ is supported by the experimental data.
 The considered sum rule, under the $SM$ conditions, is reported to be of order of $\sim2\%$ but significantly below zero. At the same time, it is very weakly constrained when the $NP$ is allowed.

The article is structured in the following manner. In Section~\ref{sect:B2Kpi_puzzle}, the current status of the so-called ``$B\to K\pi$ puzzle'' is sketched. An $SU(3)$-based sum rule $\Delta^{SR} \approx 0$ is introduced. Next, in Section~\ref{sect:diagrams}, the diagrammatic approach to $B\to PP'$ decays ($P, P'$ - pseudoscalars) is presented providing the parameterisation used later in the fit. The following Section~\ref{sect:SU3} describes how the $SU(3)$ assumption is applied. Section~\ref{sect:data} contains the list of the experimental inputs. The results of the fits to the newest data for the $SM$ and the $NP$ scenarios are presented in Section~\ref{sect:fits}. At the end the article is summarised and the prospects for the future are given.
\section{$B\to K\pi$ puzzle}
\label{sect:B2Kpi_puzzle}
The $B\to K\pi$ decays, despite being dominated by the QCD loop topologies in the SM, are potentially sensitive to the NP expected at the loop-level transitions. In particular, the EW penguin diagrams have significant impact on the total amplitudes \cite{Buras:2003yc,Buras:2003dj,Buras:2004ub,Baek:2004rp}. 
The $CP$ symmetry breaking appears as a result of the interference between the tree- and loop-level contributions. It is a consequence  of the nonzero strong and weak phase differences between the two interfering terms.

The $(B^0\to K^+\pi^-$)\footnote{The inclusion of charge-conjugated processes is implied throughout this Letter unless stated otherwise.} measurement in B-factory experiments resulted in the first observation of the $CP$ symmetry violation in the $B$ system \cite{BaBar:2004gyj,Belle:2004nch}.
When the full set of results, consisting of the $CP$ asymmetries and $BFs$, had been provided  by the B factories, Tevatron and LHCb,
the ``$B\to K\pi$ puzzle'' has been stated. The experimental input, at that time, was given by the following measurements:   
$B^0\to K^+\pi^-$~\cite{BaBar:2012fgk,Belle:2012dmz,CDF:2014pzb,LHCb:2018pff},   
$B^+\to K^+\pi^0$~\cite{Belle:2012dmz,BaBar:2007uoe},    
$B^0\to K^0\pi^0$~\cite{BaBar:2008ucf,Belle:2008kbm} and    
$B^0\to K^0\pi^0$~\cite{Belle:2012dmz,BaBar:2006enb,LHCb:2013vip}.
All of the results given above are used in the study described in~\cite{Fleischer:2017vrb}. 

The puzzle is based on the quantity~\eqref{eq:DA_CP},
which should vanish under the isospin symmetry~\cite{Buras:2003yc,Buras:2003dj,Buras:2004ub,Baek:2004rp,Baek:2007yy,Baek:2009pa,Fleischer:2017vrb,Gronau:2005kz,Ciuchini:2008eh,Beaudry:2017gtw}. On the other hand, the present experimental world average amounts to
\begin{equation}
    \Delta A_{CP}(K\pi) = 0.114\pm0.014
    \label{eq:DA_CP_exp}
\end{equation}
which is over $8\sigma$ above zero. Moreover, the SM prediction obtained with the use of the QCD factorisation~(QCDF) approach results in \newline $\Delta A_{CP}(K\pi) = (0.018_{-0.032}^{+0.041})$~\cite{crivellin:2019isj}, which is also in disagreement with the experimental outcome.  

It is noteworthy that, in the above, the isospin requirement is not applied in a fully correct manner. It should be imposed on the amplitudes, not the asymmetries. 

This analysis includes as well the study of the more fundamental sum rule based on the rate asymmetries. It can be expressed in terms of the $CP$ asymmetries and $BFs$~\cite{Gronau:1998fn}:
\begin{align}
    \Delta&_{SR} =A^{\pi^+K^-}-A^{\pi^0K^0}\frac{2BF(\pi^0K^0)}{BF(\pi^+K^-)}\nonumber\\
    &+\left[A^{\pi^+K^0}\frac{BF(\pi^+K^0)}{BF(\pi^+K^-)}-A^{\pi^0K^+}\frac{2BF(\pi^0K^+)}{BF(\pi^+K^-)}\right]\frac{\tau_{B_d}}{\tau_{B^{\pm}}}\nonumber\\
    &~~\approx0,
    \label{eq:sumrule}
\end{align}
where $BF$ stands for the branching fraction and $\tau$ is the $B$ meson average decay time. This quantity is almost perfectly equal to zero when applying the parameterisation given in Section~\ref{subsect:par1}.

The analysis presented here, is not only an update of~\cite{Fleischer:2017vrb} with the use of the newest experimental results. It adds to the metodology the $SU(3)$ decomposition approach~\cite{Gronau:1998fn} and provides the profile plot for the parameters of interest (sensitive to the $NP$). A more detailed study of the sum rule~\cite{Gronau:2005kz} is carried out.

To summarise, the SM is tested by searching for a footprint of the NP in the EW penguin amplitudes. Several assumptions are made. 
The $B\to\pi K$, $B\to\pi\pi$ and $B\to KK$ processes are fully described by graphs listed in Fig.~\ref{fig:diagrams}, where each graph represents a topology - sum of all multi-loop diagrams represented by a single complex parameter. 
The W-exchange, annihilation and penguin annihilation topologies (\emph{c.f.} Section~\ref{sect:diagrams}) can be neglected.
The isospin relations for amplitudes are fulfilled.
The $SU(3)$-based relations hold between the $B\to\pi K$ and $B\to\pi\pi$ as well as $B\to\pi K$ and $B\to KK$ decays within a margin for $SU(3)$ violation (Section~\ref{sect:SU3}). The $SU(3)$ decomposition of the effective hamiltonian leading to the relation of the penguin and tree amplitudes is valid.
Using these assumptions a model prediction is established. In the case of its inconsistency with the experimental outcomes one (or more) of the above premises is false or the SM is incomplete and there is a hint towards NP. The details of the method are presented in the following sections.

\section{The Feynman topology decomposition of $B\to PP'$ decays}
\label{sect:diagrams}
\begin{figure}
    \centering
    \begin{tabular}{ccc}
        \begin{subfigure}{0.14\textwidth}
            \includegraphics[width=\textwidth]{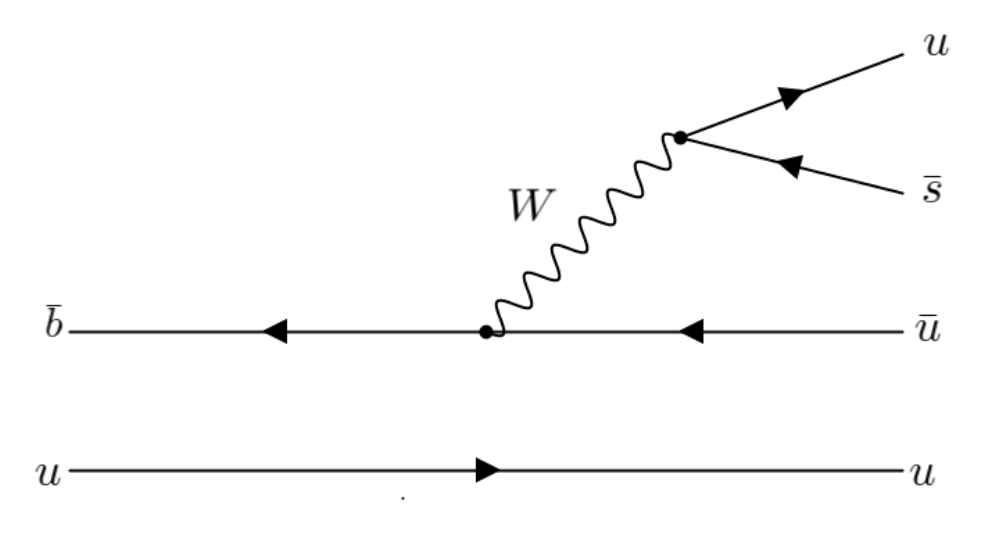} 
            \subcaption{tree colour-allowed $\mathcal{T'}$}
            \label{fig:diagrams_tree_ca}
        \end{subfigure}&       
        \begin{subfigure}{0.14\textwidth}
            \includegraphics[width=\textwidth]{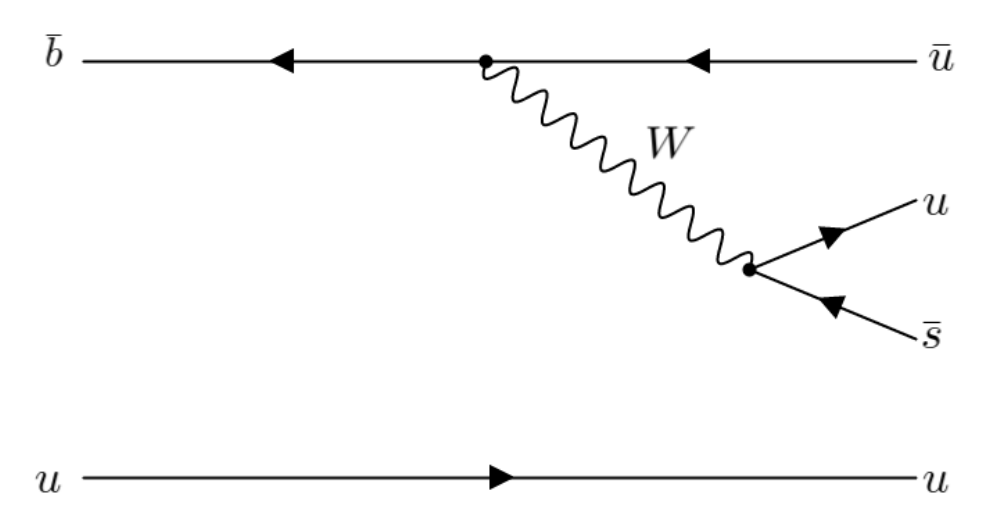}
            \subcaption{tree colour-suppressed \\\centering $\mathcal{C'}$}
            \label{fig:diagrams_tree_cs}
        \end{subfigure}&
        \begin{subfigure}{0.14\textwidth} 
            \includegraphics[width=\textwidth]{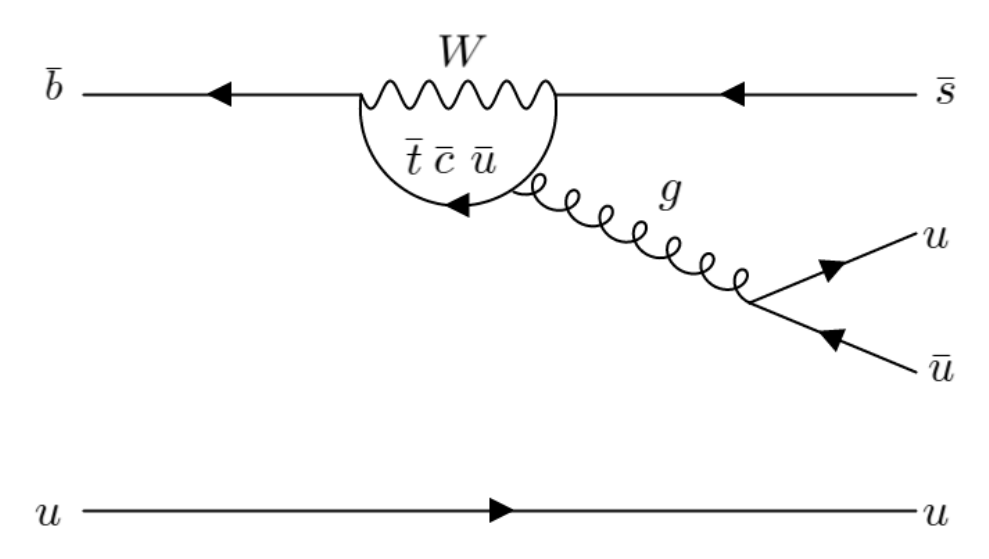}  
            \subcaption{QCD penguin~$\mathcal{P}_{q}'$ \\\centering $q=t,u,c$}
            \label{fig:diagrams_QCD}
        \end{subfigure}\\
        \begin{subfigure}{0.14\textwidth}
            \includegraphics[width=\textwidth]{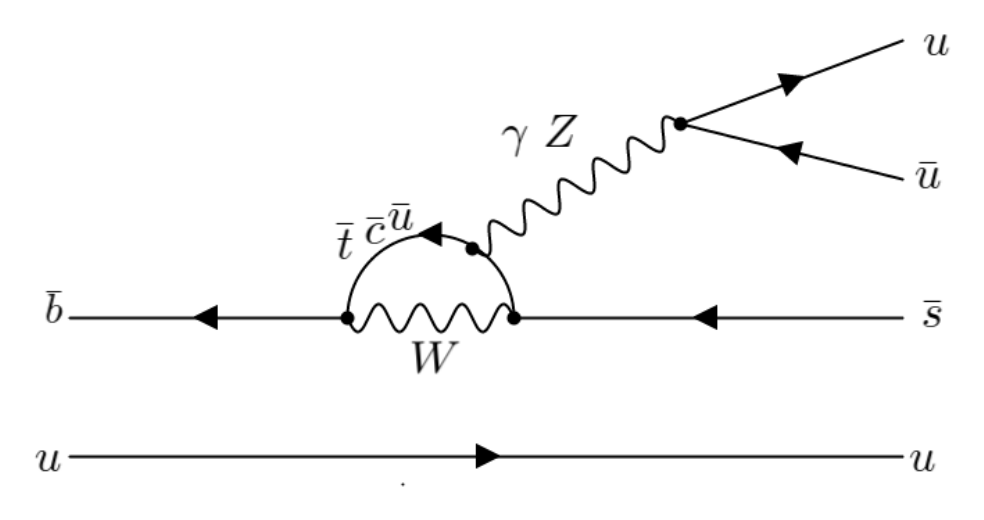}
            \subcaption{EW penguin colour-allowed $\mathcal{P}_{EW}'$}
            \label{fig:diagrams_EW_ca}
        \end{subfigure}&
        \begin{subfigure}{0.14\textwidth}
            \includegraphics[width=\textwidth]{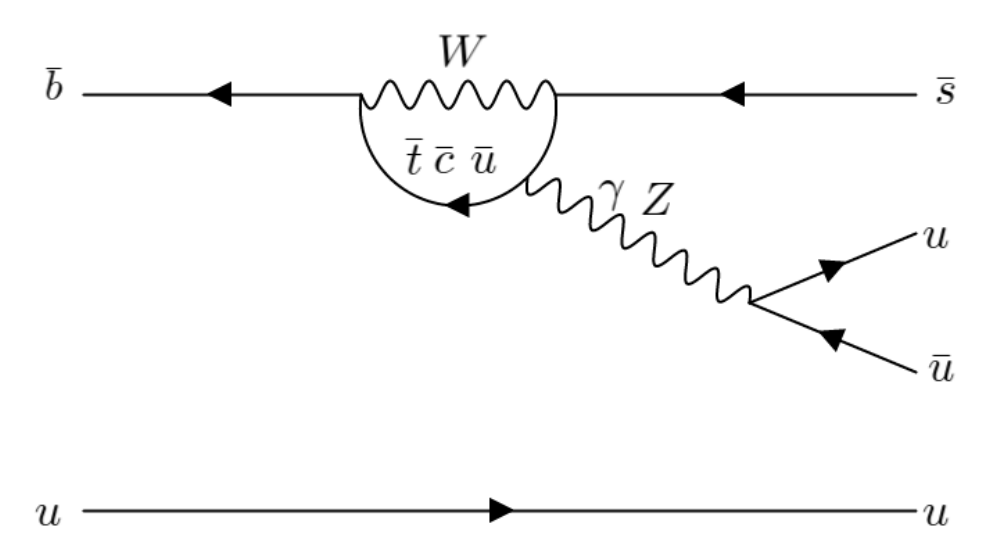} 
            \subcaption{EW penguin colour-suppressed $\mathcal{P}_{EW}^C{'}$}
            \label{fig:diagrams_EW_cs}
        \end{subfigure}& 
        \begin{subfigure}{0.14\textwidth}
            \includegraphics[width=\textwidth]{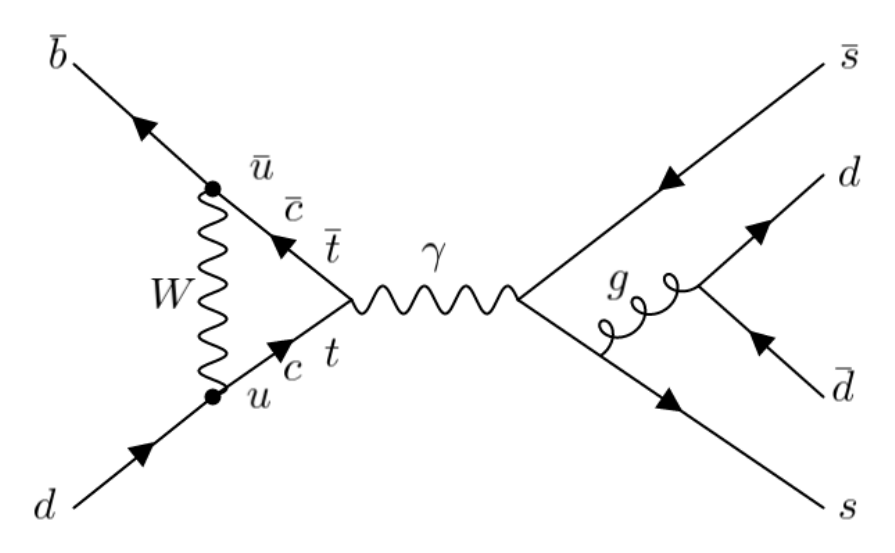}
            \subcaption{penguin annihilation \\\centering $\mathcal{PA'}$}
            \label{fig:diagrams_penguin_annihilation}
        \end{subfigure}\\
                \begin{subfigure}{0.14\textwidth}
            \includegraphics[width=\textwidth]{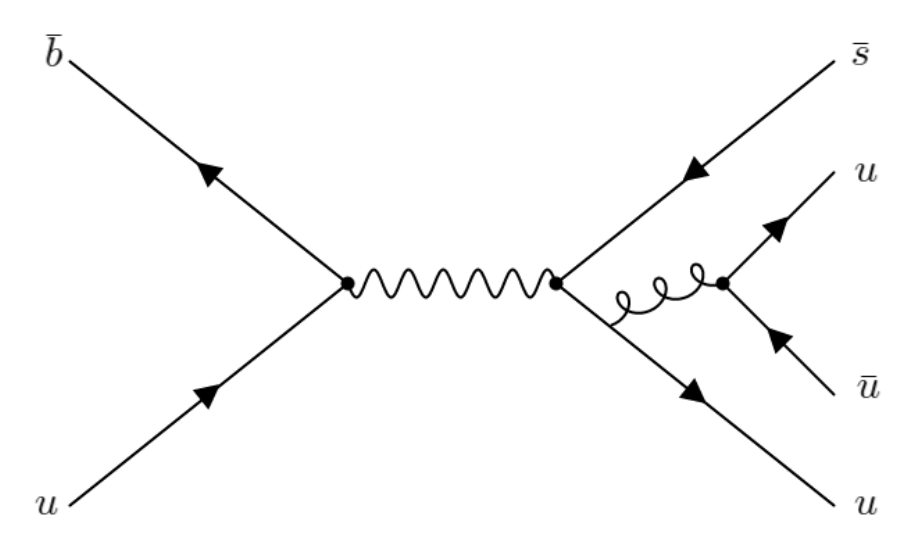}
            \subcaption{annihilation $\mathcal{A'}$}
            \label{fig:diagrams_annihilation}
        \end{subfigure}&
        \begin{subfigure}{0.14\textwidth}
            \includegraphics[width=\textwidth]{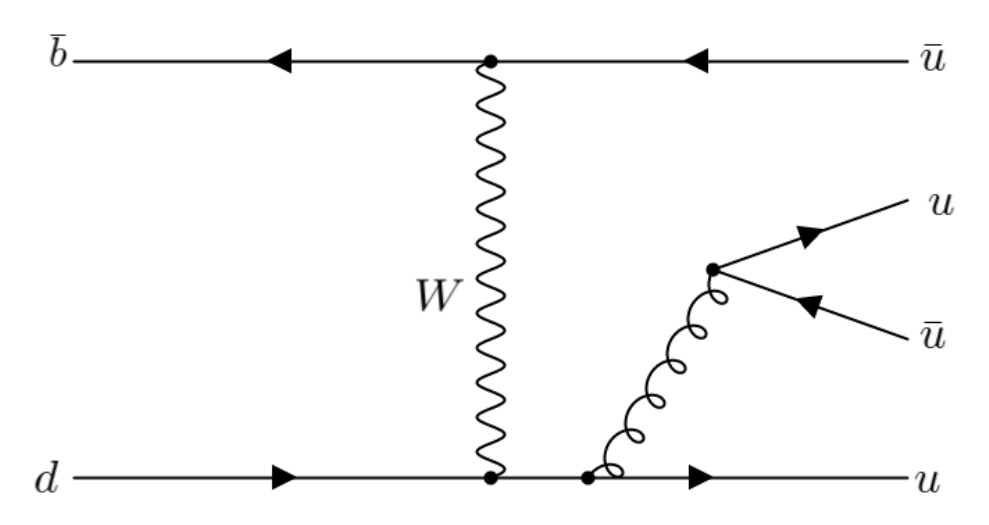}
            \subcaption{W-exchange $\mathcal{E}$}
            \label{fig:diagrams_ex}
        \end{subfigure}&\\
    \end{tabular}
    \caption{The Feynman graphs representing the $B\to PP'$ decays. As an example in panels~(a-g) $B^+\to K^+\pi^0$ graphs are given. In panel~(h) W-exchange topology of $B^0\to\pi^0\pi^0$ is presented. Diagrams representing $B\to \pi\pi$ have the same topologies and are obtained by substituting of the spectator quark or changing the transition $b\to s$ to $b\to d$. In this notation, the prime ($'$) symbol indicates the $b\to s$ decays.}
    \label{fig:diagrams}
\end{figure}

This section describes the diagrammatic approach to represent the $B\to PP'$ decays. 
Feynman graphs representing decay topologies of $B\to PP'$ decays are shown in Fig.~\ref{fig:diagrams}. 
The tree amplitudes dominate the $B\to\pi\pi$ reaction while the $EW$ penguin contribution is small there. In the $B\to K\pi$ decays tree topologies are suppressed by CKM matrix factors and the impact of $EW$ penguins is measurable. In both cases, QCD penguin amplitudes play a major role, while the $W$-exchange, penguin annihilation and annihilation contributions can be neglected. As noted in~\cite{Hernandez:1994re} $\mathcal{E}^({'}^)$ and $\mathcal{A}^({'}^)$\footnote{Throughout this paper the prime symbol ($'$) represents the $b\to s$ transition as opposed to $b\to d$ transitions.} are helicity suppressed by $m_{u,d,s}/m_B$ as $B$ is a pseudoscalar. Moreover, the contribution of all exchange and annihilation amplitudes, like $\mathcal{E}^({'}^)$,  $\mathcal{A}^({'}^)$ and $\mathcal{PA}^({'}^)$, is weakened by the factor of the $B$-meson wave function at the origin - $f_B/m_B \lesssim 0.05$ with respect to the tree topologies. They are, however, presented here for completeness. 

The $B\to\pi\pi$ decomposition reads:
\begin{subequations}
\label{eq:pipi_1}
\begin{flalign}
    \sqrt{2}\mathcal{A}(B^+\to \pi^0\pi^+)&=-\lambda^3A R_b\Bigg[e^{i\gamma} (\mathcal{T}+\mathcal{C}+\mathcal{A})\nonumber\\&+e^{-i\beta}(P_{EW}+P_{EW}^C)\Bigg], \label{eq:pi0pip_1}\\
    \mathcal{A}(B^0\to\pi^-\pi^+)&=-\lambda^3 AR_b \nonumber \\\times\Bigg[e^{i\gamma}(\mathcal{T}-\mathcal{P}_{tu}&+\mathcal{E}-\mathcal{PA}_{tu}) +\frac{1}{R_b}(\mathcal{P}_{tc}+\mathcal{PA}_{tc})\Bigg],\label{eq:pimpip_1}\\
    \sqrt{2}\mathcal{A}(B^0\to\pi^0\pi^0)&=\lambda^3AR_b\Bigg[\frac{1}{R_b}(\mathcal{P}_{tc}+\mathcal{PA}_{tc})\nonumber \\
    -e^{i\gamma}(\mathcal{C}+\mathcal{P}_{tu}&-\mathcal{E}+\mathcal{PA}_{tu})+e^{-i\beta}(P_{EW}+P_{EW}^C)\Bigg]. \label{eq:pi0pi0_1}
\end{flalign}
\end{subequations}
Here, $\mathcal{T}$ and $\mathcal{C}$ represent the tree colour-allowed and tree colour-suppressed amplitudes respectively. Similarly, $\mathcal{P}_{EW}$ and $\mathcal{P}_{EW}^C$ stand for the colour-allowed and colour-suppressed $EW$ penguin topologies. 
The QCD penguin contribution is parameterised by $\mathcal{P}_{tc} = \mathcal{P}_t-\mathcal{P}_c$ and $\mathcal{P}_{tu} = \mathcal{P}_t-\mathcal{P}_u$. Here, $P_q$ denotes the QCD penguin amplitude goverened by the quark $q=t, c, u$  and  $\gamma= arg\left[-\frac{V_{ud}V_{ub}^*}{V_{cd}V_{cb}^*}\right]$ is a relative CKM weak phase.
The penguin annihilation contributions are labelled with $\mathcal{PA}$, while $\mathcal{E}$ and $\mathcal{A}$ stand for the $W$-exchange and annihilation amplitude respectively. The weak phase $\beta$ is one of the CKM unitarity triangle angles - $\beta = arg\left[-\frac{V_{cd}V_{cb}^*}{V_{td}V_{tb}^*}\right]$. 
The strong phases are not shown explicitly, they are contained within the complex parameters representing each amplitude. They are undestood as $CP$-even, while the weak phases as $CP$-odd ( changing sign when anti-B mesons are considered).
The Wolfenstein parameters of the CKM matrix amount to $\lambda\equiv|V_{us}|=0.22500^{+0.00024}_ {-0.00022}$ and $A\equiv|V_{ub}|/\lambda^2=0.8132^{+0.0119}_{-0.0060}$ \cite{Wolfenstein:1983yz,Buras:1994ec,CKMfitter}.
Lastly, the $R_b$ factor measures the side of the UT triangle and is given by $R_b\equiv\left(1-\frac{\lambda^2}{2}\right)\frac{1}{\lambda}\left|\frac{V_{ub}}{V_{cb}}\right| = 0.381\pm0.011$
as reported in \cite{CKMfitter}.

One important fact should be stressed at this point. In Eqs.~\eqref{eq:pipi_1} it is implicitly assumed, that interchanging the spectator quark $u\leftrightarrow d$ does not change the relative magnitudes or phases between the amplitudes. In other words, the isospin relation of the amplitudes, $\sqrt{2}\mathcal{A}(B^+\to \pi^+\pi^0) = \mathcal{A}(B^0\to\pi^-\pi^+)+\sqrt{2}\mathcal{A}(B^0\to\pi^0\pi^0)$, is implicitly fulfilled within this parameterisation. This application of the isospin symmetry, at the level of amplitudes, is held throughout this paper.

It should be also noted here, that the proposed analysis operates on asymmetries and ratios of $BFs$ and therefore any normalisation terms in the amplitudes cancel out, although the relative normalisation between $b\to d$ and $b\to s$ has to be taken into account.  In this analysis, this ratio is a free parameter.

In the following, the parameterisation of the $B\to\pi\pi$, $B\to K\pi$ and $B\to KK$ amplitudes is given. It is based on the solutions proposed in~\cite{Buras:2003yc,Buras:2003dj} - polar representation of the complex variables - and representation by real and imaginary parts applied, for example, in the analysis~\cite{Hofer:2010ee}. 
The former brings insight into the possible $NP$ contribution, namely: $qe^{i\phi}$ while the latter is more suitable for fitting. 

\subsection{Parameterisation}
\label{subsect:par1}
The following notation is chosen:
\begin{subequations}
\label{eq:B2pipi_def}
\begin{align}
    	\Tilde{T} &= \lambda^3AR_b(\mathcal{T}-\mathcal{P}_{tu}),\\
	\Tilde{C} &=\lambda^3AR_b(\mathcal{C}+\mathcal{P}_{tu}),\\
	\Tilde{P}&=\lambda^3A(\mathcal{P}_t-\mathcal{P}_c) = \lambda^3A\mathcal{P}_{tc},\\
	r_T &= \frac{\Tilde{T}}{P}~~~~r_C = \frac{\Tilde{C}}{P}.
\end{align}
\end{subequations}
Applying this to Eqs.~\eqref{eq:pipi_1} and ignoring the annihilation amplitudes $\mathcal{A}$ gives
\begin{subequations}
    \label{eq:B2pipi}
\begin{flalign}
        \sqrt{2}\mathcal{A}(B^+\to \pi^0\pi^+)=&-P\Big[e^{i\gamma} (r_T+r_C)\nonumber \\&+\tilde{q}e^{i(-\beta+\tilde{\phi}+\tilde{\omega})}(r_T+r_C)\Big], \label{eq:pi0pip_short_1}\\
    \mathcal{A}(B^0\to\pi^-\pi^+)=&-P(1-r_Te^{i\gamma}),\label{eq:pimpip_short_1}\\
        \sqrt{2}\mathcal{A}(B^0\to\pi^0\pi^0)=&~P\Big[1-e^{i\gamma}r_C\nonumber \\&+\tilde{q}e^{i(-\beta+\tilde{\phi}+\tilde{\omega})}(r_T+r_C)\Big]. \label{eq:pi0pi0_short_1}
\end{flalign}
\end{subequations}
In the above the $EW$ penguin contribution $\tilde{q}e^{i(\tilde{\phi}+\tilde{\omega})} = (P_{EW}+P_{EW}^C)/(T+C)$ is small:
\begin{align}
        \Tilde{q}&\equiv\left|\frac{\mathcal{P}_{EW}+\mathcal{P}_{EW}^C}{\mathcal{T+C}}\right|\sim\frac{3}{2}\frac{C_9(\mu)+C_{10}(\mu)}{C_1(\mu)+C_2(\mu)}\left|\frac{V_{td}}{V_{ub}}\right|\nonumber \\&\approx1.3\times 10^{-2}\left|\frac{V_{td}}{V_{ub}}\right|\approx 3\times 10^{-2}, \label{eq:tilde_q}
\end{align}
where $C_i(\mu)$ are the Wilson coefficients obtained at the given scale $\mu$. Their values can be obtained in perturbative QCD for $\mu=m_b$~\cite{Buchalla:1995vs}.  In~\cite{Buras:2004ub,Fleischer:2018bld} the impact of $EW$ penguin amplitudes on $b\to d$  transitions (Eq.~\eqref{eq:tilde_q}) is omitted as 
their effect on the determination of the hadronic parameters - magnitudes and strong phases of $T$ and $C$ - is negligible. In this analysis, however, this contribution is kept in the fit.

Basing on the expressions  presented in~\cite{Fleischer:2018bld} and using a suitable notation the $B\to K\pi$ amplitudes are expressed:
    \begin{subequations}
    \label{eq:B2Kpi}
    \begin{alignat}{2}
            &A(B^+&&\to K^0\pi^+)= -P'\Bigg[1+r'_{\rho}e^{i\gamma}\nonumber \\& &&-\frac{1}{3}\hat{a}'_C\boldsymbol{qe^{i\omega}e^{i\phi}}(r'_T+r'_C)\Bigg],\\
            \sqrt{2}&A(B^+&&\to K^+\pi^0)=P'\Bigg[1+r'_{\rho}e^{i\gamma}-\Big\{e^{i\gamma}\nonumber \\& &&-\Big(1-\frac{1}{3}\hat{a}'_C\Big)\boldsymbol{qe^{i\omega}e^{i\phi}}\Big\}(r'_T+r'_C)\Bigg],\\
            &A(B_d^0&&\to K^+\pi^-)=P'\Bigg[1+r'_{\rho}e^{i\gamma}\nonumber \\& &&+\frac{2}{3}a'_C\boldsymbol{qe^{i\omega}e^{i\phi}}(r'_T+r'_C)-r'_Te^{i\gamma}\Bigg],\\
            \sqrt{2}&A(B_d^0&&\to K^0\pi^0)=-P'\Bigg[1+r'_{\rho}e^{i\gamma}-r'_Te^{i\gamma}\nonumber \\& &&+\Big\{e^{i\gamma}-\Big(1-\frac{2}{3}a'_C\Big)\boldsymbol{qe^{i\omega}e^{i\phi}}\Big\}(r'_T+r'_C)\Bigg],
    \end{alignat}
    \end{subequations}
    where
    \begin{subequations}
    \begin{align}
      P'&\equiv\frac{\lambda^3A}{\sqrt{\epsilon}}(\mathcal{P}'_t-\mathcal{P}'_c),\label{eq:P_prime}\\
      r_{\rho}&\equiv\epsilon R_b\left[\frac{\mathcal{P}'_t-\mathcal{\tilde{P}}'_u-\mathcal{A}'}{\mathcal{P}'_t-\mathcal{P}'_c}\right],\\
      a'_C&\approx\hat{a}'_C\equiv\frac{\mathcal{\hat{P}}_{EW}'^C}{\mathcal{\hat{P}}'_{EW}+\mathcal{\hat{P}}_{EW}'^C},&&\label{eq:a_C}\\
      r'_T&\equiv\epsilon R_b\left[\frac{\mathcal{T'}}{\mathcal{P}'_t-\mathcal{P}'_c}\right]\equiv\frac{\hat{T}'}{P'},\\
      r'_C&\equiv\epsilon R_b\left[\frac{\mathcal{C'}}{\mathcal{P}'_t-\mathcal{P}'_c}\right]\equiv\frac{\hat{C}'}{P'}
    \end{align}
    \end{subequations}
    and (\cite{Fleischer:2018bld}):
    \begin{equation}
        \epsilon\equiv\frac{\lambda^2}{1-\lambda^2} =  0.0535 \pm 0.0002.
    \end{equation}
    Analogous equations for the $B\to KK$ case are given in the next section.
    
    In the above Eq.~\eqref{eq:a_C} the $\hat{a}_C$ represents the contribution of the $EW$ colour-suppressed penguin to $B^+$ decays as opposed to $a_C$ which stands for the same contribution in $B^0$ decays. Here the difference in the spectator quark $u\leftrightarrow d$ is given explicitly and the equality between the two amplitudes is a result of the isospin symmetry application.
    
    In the formulae for the $B\to K\pi$ decay amplitudes - \eqref{eq:B2Kpi} - the amount of the contribution of the  $EW$ penguin topologies is highlighted in bold:
    \begin{equation}
        qe^{i\phi}e^{i\omega}= -\frac{\mathcal{P}_{EW}'+\mathcal{P}_{EW}'^C}{\mathcal{T}'+\mathcal{C}'},
    \end{equation}
    where $\phi$ is the weak phase ($CP$-odd phase) and $\omega$ is the strong phase ($CP$-even phase). In the $SM$ $\phi=0$, since applying the operator expansion of the effective hamiltionian, omitting the small Wilson coefficients $c_7$ and $c_8$ and using the Fierz identities one notices a linear dependence between the tree and $EW$ penguin operators (\cite{Buchalla:1995vs}). To constrain $q$ and $\omega$ the isospin decomposition is used. The full derivation, given in~\cite{Neubert:1998pt}, uses the following SU(3)($U$-spin) relation:
    \begin{align}
        &&\mathcal{T}'+\mathcal{C}' = -\sqrt{2}\frac{V_{us}}{V_{ud}}\frac{f_K}{f_{\pi}}\mathcal{A}(B^+\to \pi^+\pi^0),
        \label{eq:SU3_pippi0}
    \end{align}
    where the factor $f_K /f_{\pi} = 1.22 \pm 0.01$ accounts for the leading $SU(3)$-breaking corrections calculable in QCD factorisation regime. Equation~\eqref{eq:SU3_pippi0} corresponds via the $SU(3)$ relation to the $B^+\to K^+\pi^0$ channel. The argumentation leads to the following result:
    \begin{equation}
        \label{eq:q_phi_omega}
	    qe^{i\phi}e^{i\omega} = -\frac{3}{2\lambda^2R_b}\frac{c_9(\mu)+c_{10}(\mu)}{c_1(\mu)+c_2(\mu)} = (0.64\pm0.05)R_q,
    \end{equation}
    which can be compared to the older value reported in~\cite{Fleischer:2018bld}. With respect to the $b\to d$ transitions, Eq.~\eqref{eq:tilde_q}, q is enhanced by $\frac{1}{\lambda^2R_b}\approx50.65$. A possible deviation from the $SU(3)$ symmetry between $B^+\to \pi^0\pi^+$ and $B^+\to K^+\pi^0$ is described by $R_q=1.00\pm 0.05$. The Wilson coefficients are labelled by $c_i$. Coefficients $c_1$ and $c_2$ correspond to the tree operators, while $c_9$ and $c_{10}$ refer to the $EW$ penguin contribution. The $NP$ contribution can potentially effectively change the Wilson coefficient values and therefore $q$ and the weak phase $\phi$. In the analysis below, the strong phase $\omega=0$ is assumed. Under the described conditions $\omega$ vanishes both in the $SM$ and in the $NP$ case when $SU(3)$ is conserved.
    
    In the $SU(3)$ decomposition approach, proposed \newline in~\cite{Gronau:1998fn} and applied in the analysis described in~\cite{Baek:2004rp},  the decay amplitudes are given in terms of $SU(3)$-reduced transition matrix elements, labelled in the following way~\cite{Gronau:1998fn}:
    \begin{subequations}
    \begin{align}
    a_3 &= -\lambda^{(s)}_t\frac{1}{2}(c_1-c_2)\langle\boldsymbol{8||6||3}\rangle,\\b_3 &= \frac{3}{2}\lambda^{(s)}_u(c_9-c_{10})\langle\boldsymbol{8||6||3}\rangle,\\
    a_4 &= \frac{1}{2}\lambda^{(s)}_t(c_1+c_2)\langle\boldsymbol{8||\overline{15}||3}\rangle, \\b_4 &= \frac{1}{2}\lambda^{(s)}_u(c_9+c_{10})\langle\boldsymbol{8||\overline{15}||3}\rangle, \\
    a_5 &= \frac{1}{2}\lambda^{(s)}_t(c_1+c_2)\langle\boldsymbol{27||\overline{15}||3}\rangle,\\b_5 &=\frac{1}{2}\lambda^{(s)}_u(c_9+c_{10})\langle\boldsymbol{27||\overline{15}||3}\rangle,
    \end{align}
    \end{subequations}
    one may find the relation to the parameters given in Eqs.~\eqref{eq:B2Kpi}
    \begin{subequations}
    \begin{align}
    b_3 &= \frac{\lambda^{(s)}_u}{\lambda^{(s)}_t}\frac{c_9-c_{10}}{c_1-c_2}\frac{\sqrt{15}}{2}(C'-T')\\
    b_4 &= -\frac{1}{2}\sqrt{\frac{1}{15}}\frac{\lambda^{(s)}_u}{\lambda^{(s)}_t}\frac{c_9+c_{10}}{c_1+c_2}(C'+T')\\
    b_5 &=-\frac{3}{\sqrt{10}}\frac{\lambda^{(s)}_u}{\lambda^{(s)}_t}\frac{c_9+c_{10}}{c_1+c_2}(C'+T')
    \end{align}
    \end{subequations}
    where the exchange and annihilation contributions are neglected. Finally, the connection between the $EW$ penguin and the tree amplitudes is reached
    \begin{align}
    \label{eq:EWP_Kpi_SU3}
	    P'_{EW}+P'_{EW}{^C} &= -\sqrt{\frac{5}{2}}b_5 = \frac{3}{2}\frac{\lambda^{(s)}_u}{\lambda^{(s)}_t}\frac{c_9+c_{10}}{c_1+c_2}(T'+C')\nonumber\\& = -qe^{i\phi}(T'+C')
    \end{align}
    as in the previous approach.
    Moreover, the pure $EW$ color-suppressed penguin contribution can be expressed by tree topologies:
    \begin{align}
            &P'_{EW}{^C} =\frac{1}{2}\sqrt{\frac{3}{5}}b_3
            +\frac{3}{2}\sqrt{\frac{3}{5}}b_4-\frac{3}{2}\sqrt{\frac{2}{5}}b_5\nonumber\\
                         &~~=-\frac{1}{2}\Bigg[-\frac{3}{2}\frac{\lambda_u^{(s)}}{\lambda_t^{(s)}}\frac{c_9-c_{10}}{c_1-c_2}(T'-C')\nonumber\\&~~~~~+\frac{3}{2}\frac{\lambda_u^{(s)}}{\lambda_t^{(s)}}\frac{c_9+c_{10}}{c_1+c_2}(T'+C')\Bigg]\nonumber \\&~~=-qe^{i\phi}C'.
        \label{eq:EWPC_SU3}
    \end{align}
    In the last line, numerical approximation $\frac{c_1-c_2}{c_9-c_{10}}\approx\frac{c_1+c_2}{c_9+c_{10}}$ is used. Here, it is assumed that the $NP$ represented by $qe^{i\phi}$ enters evenly in all the reduced transition matrix elements.
    As a consequence of relations \eqref{eq:EWP_Kpi_SU3} and \eqref{eq:EWPC_SU3} it follows: 
    \begin{align}
        P'_{EW} = -qe^{i\phi}T'.
    \end{align}
    Moreover, the $EW$ penguin amplitudes in $B\to\pi\pi$ decays may be related to the tree amplitudes:
    \begin{align}
        \label{eq:EWPpipi}
	    P_{EW}+P_{EW}^C &=\-\frac{1}{2}\frac{\sqrt{5}}{2}b_5= -\lambda^2qe^{i\phi}(T+C), 
    \end{align}
where the influence of $NP$ is again parameterised by $q$ and $\phi$.

\section{$SU(3)$ constraints.}
\label{sect:SU3}
The $SU(3)$ symmetry is applied in four different ways in this analysis. The first one is the isospin symmetry included in the amplitude decomposition and parameterisation of the considered decays given in Eqs.~\eqref{eq:pipi_1} and~\eqref{eq:B2Kpi}. It is intrinsically assumed that the same topologies in $B^0$ and $B^+$ decays are represented by the same parameters. 

The second way the $SU(3)$ is exploited is by the decomposition of the effective hamiltonian given in the previous section. The relations \eqref{eq:EWP_Kpi_SU3}, \eqref{eq:EWPC_SU3} and \eqref{eq:EWPpipi} are used to reduce the number of parameters in the fit.

Thirdly, the application of the $SU(3)$ symmetry is concerned  with the  $s\longleftrightarrow d$ interchange ($U$-spin symmetry). Here, the relation between amplitudes of $B\to \pi\pi$ and $B\to K\pi$ decays is taken into consideration. It is a well-known fact that $SU(3)$ symmetry is broken here due to the difference between the $s$ and $d$ quark masses. A loose relation, however, can be applied, as described in \cite{Fleischer:2018bld}. Magnitudes of tree amplitudes are related by
\begin{align}
    R_{T+C} = \frac{|\mathcal{T}'+\mathcal{C}'|}{|\mathcal{T}+\mathcal{C}|} = \frac{|r'_T+r'_C|}{|\epsilon(r_T+r_C)|} = 1.2 \pm 0.2.
    \label{eq:R_TC}
\end{align}
The central value is obtained in the factorisation framework. The uncertainty follows from a conservative approach by taking into account non-factorisable $SU(3)$-breaking effects as large as 100\% \cite{Fleischer:2008wb}. Following the arguments given in \cite{Fleischer:2018bld}, considering corrections of $\sim20\%$ to the $SU(3)$ symmetry, loose constraints on tree amplitude phases can be imposed:
\begin{subequations}
\begin{align}
    arg(\mathcal{T}')-arg(\mathcal{T}) &= (0\pm20)^{\circ},\\
    arg(\mathcal{C}')-arg(\mathcal{C}) &= (0\pm20)^{\circ}.
\end{align}
\label{eq:SU3_phases}
\end{subequations}
The above relations, Eqs.\eqref{eq:R_TC} and \eqref{eq:SU3_phases} are applied as extra constraints to the fit bringing relation between $b\to d$ and $b\to s$ transitions. 

The fourth way is a relation between the $B\to K\pi$ and $B\to KK$ decays. The $B\to KK $ amplitudes in the diagrammatic approach are given by
\begin{subequations}
    \begin{flalign}
    \mathcal{A}(B^+\to K^+K^0)=& ~P^{KK} \Big(1+r_{\rho}^{KK}e^{i\gamma}\nonumber \\&+\frac{1}{3}a^{KK}_C(r_T+r_C)qe^{i(\omega+\phi)}\Big),\\
    \mathcal{A}(B^0\to K^0K^0)=& ~P^{KK} \Big(1+r_{\rho}^{KK}e^{i\gamma}\nonumber\\&+\frac{1}{3}a^{KK}_C(r_T+r_C)qe^{i(\omega+\phi)}\Big).
    \end{flalign}
\end{subequations}
In this analysis, it is assumed that
\begin{align}
    \label{eq:KK_rel}
    r_{\rho}^{KK} = r'_{\rho}, ~~~~~~~~ a_C^{KK} = a_C,
\end{align}
while $P^{KK}$ is left free independently of $P' (K\pi)$. The relation \eqref{eq:KK_rel} is set to hold strictly, as both $\rho_c^{KK}$ and $a_C^{KK}$ express amplitudes relative to $P^{KK}$ similarly as $r'_{\rho}$ and $a_C$ are relative to $P'$ in the $B\to K\pi$ reaction.

\section{Experimental data}
\label{sect:data}
The data consists of up-to-date results of the $CP$ asymmetry and the $BF$ measurements. Most results are listed in the Particle Data Group 2022 review~\cite{ParticleDataGroup:2022pth}, where the averages and their uncertainties can be found. When a newer measurement is included its error is composed of the statistical and the systematic uncertainties $\sigma_{exp} = \sqrt{\sigma^2_{stat}+\sigma^2_{syst}}$. The new mean is obtained as the weighted average of the current value and the new result. The $BFs$ have to be phase-space corrected according to the masses of the mesons involved (see~\ref{app:corr}).

In Table~\ref{tab:data_pipi} the set of  $B\to\pi\pi$ data is presented. It is noteworthy that only the relative $BF$ values are relevant for the fit. The $CP$ asymmetries and the $BFs$ of the $B\to K\pi$ decays are shown in Table~\ref{tab:data_Kpi}. In addition, the $B$ decays to two kaons, $B\to K^+K^0$ and $B\to K^0K^0$, are used to constrain the $r_{\rho}$ amplitude. The data is given in Table~\ref{tab:data_KK}. 

\begin{table}[tbh]
\centering
\begin{tabular}{|c|c|c|}
\hline
      Observable & experimental value & source\\\hline
       $A^{\pi^+\pi^-}$   & $0.314\pm0.030$ & \makecell{\cite{ParticleDataGroup:2022pth}\\\cite{Belle-II:2023ksq}} \\\hline 
       $A^{\pi^+\pi^0}$   & $0.01\pm0.04$& \makecell{\cite{ParticleDataGroup:2022pth}\\ \cite{Belle-II:2023ksq}}\\\hline
       $A^{\pi^0\pi^0}$   & $0.30\pm0.20$ & \makecell{\cite{ParticleDataGroup:2022pth}\\\cite{Belle-II:2023cbc}} \\\hline
       $S^{\pi^+\pi^-}$   & $-0.670\pm0.030$ &\cite{ParticleDataGroup:2022pth} \\\hline
	$BF(\pi^+\pi^-)$ & $(5.35\pm0.16)10^{-6}$ & \makecell{\cite{ParticleDataGroup:2022pth}\\\cite{Belle-II:2023ksq}} \\\hline
        $BF(\pi^+\pi^0)$      & $(5.30\pm0.38)10^{-6}$ & \makecell{\cite{ParticleDataGroup:2022pth}\\\cite{Belle-II:2023ksq}} \\\hline
        $BF(\pi^0\pi^0)$      & $(1.51\pm0.21)10^{-6}$ & \makecell{\cite{ParticleDataGroup:2022pth}\\\cite{Belle-II:2023cbc}} \\ \hline
\end{tabular}\\%
\caption{\label{tab:data_pipi}$B\to\pi\pi$ experimental data}
\end{table}
The time-dependent $CP$ asymmetry, $S^{hh}$, is defined by
\begin{equation}
    A^{hh}_{CP}(t)(B\to hh) = A^{hh}\cos{\Delta m_dt}+S^{hh}\sin{\Delta m_dt}.
\end{equation}
\begin{table}[tbh]
\centering
\begin{tabular}{|c|c|c|}
\hline
      Observable & experimental value & source \\\hline
	$A^{\pi^+K^-}$   & $-0.0831\pm0.0032$ & \makecell{\cite{ParticleDataGroup:2022pth}\\\cite{Belle-II:2023ksq}}\\\hline
	$A^{\pi^+K^0}$   & $-0.029\pm0.014$& \makecell{\cite{ParticleDataGroup:2022pth}\\\cite{Belle-II:2023ksq}}\\\hline
	$A^{\pi^0K^+}$   & $0.027\pm0.012$ & \makecell{\cite{ParticleDataGroup:2022pth}\\\cite{Belle-II:2023ksq}} \\\hline
	$A^{\pi^0K^0}$   & $-0.051\pm0.091$ & \makecell{\cite{ParticleDataGroup:2022pth}\\\cite{Belle-II:2023ksq}} \\\hline
	$S^{\pi^0K^0}$   & $0.64\pm0.14$ & \makecell{\cite{ParticleDataGroup:2022pth}\\\cite{Belle-II:2023grc}} \\\hline
	$BF(\pi^+K^-)$ & $(1.99\pm0.05)10^{-5}$ &  \makecell{\cite{ParticleDataGroup:2022pth}\\\cite{Belle-II:2023ksq}}\\\hline
	$BF(\pi^+K^0)$      & $(2.39\pm0.07)10^{-5}$ &  \makecell{\cite{ParticleDataGroup:2022pth}\\\cite{Belle-II:2023ksq}}\\\hline
	$BF(\pi^0K^+)$      & $(1.32\pm0.05)10^{-5}$ &  \makecell{\cite{ParticleDataGroup:2022pth}\\\cite{Belle-II:2023ksq}}\\\hline
	$BF(\pi^0K^0)$      & $(10.1\pm0.5)10^{-6}$ &  \makecell{\cite{ParticleDataGroup:2022pth}\\\cite{Belle-II:2023ksq}}\\\hline
\end{tabular}\\%
\caption{\label{tab:data_Kpi}$B\to K\pi$ experimental data}
\end{table}

\begin{table}[tbh]
\centering
\begin{tabular}{|c|c|c|}
\hline
      Observable & experimental value & source \\\hline
	$A^{K^+K^0}$ &  $0.04 \pm 0.14$ & \cite{ParticleDataGroup:2022pth}  \\\hline
	$A^{K^0K^0}$ &   $-0.6 \pm 0.7$ & \cite{ParticleDataGroup:2022pth}   \\\hline
	$BF(K^0K^+)$  &  $(1.31\pm0.17)10^{-6}$ & \cite{ParticleDataGroup:2022pth} \\\hline
	$BF(K^0K^0)$  &  $(1.21\pm0.16)10^{-6}$ & \cite{ParticleDataGroup:2022pth}\\\hline
\end{tabular}\\%
\caption{\label{tab:data_KK}$B\to KK$ experimental data}
\end{table}
Instead of using the pure $BFs$, it is better to use ratios of the form 
\begin{align}
    R^{h_1h_2}_{h'_1h'_2} = \frac{BF^{corr}(h_1h_2)}{BF^{corr}(h'_1h'_2)},
\end{align}
where $h_1, h_2, h'_1, h'_2$ stand for $K^0, K^{\pm}, \pi^0, \pi^{\pm}$. Having nine $BFs$, eight ratios are constructed. 

%
The correlation between the ratios is taken into account by calculating the covariance matrix (see~\ref{app:Cov}) and 
the contribution to the total $\chi^2$ used in the fit is given by
	\begin{align}
		\chi^2 = (R^i_{\text{exp}}-R^i_{\text{theory}})Cov^{-1}_{ij}(R^j_{\text{exp}}-R^j_{\text{theory}}),
		\label{eq:chi2_Rcov}
	\end{align}
	where $R_{\text{theory}}$ contains the parameters of the fit.
	It has been checked that the direct use of $BFs$ does not change the results. 
	
	The external CKM matrix parameters used in the fit are listed in Table~\ref{tab:data_external}.

\begin{table}[tbh]
	\centering
\begin{tabular}{|c|c|c|}
\hline
      Observable & experimental value & source \\\hline
	$\gamma$ & $(65.5\pm 1.3)^{\circ}$ & \cite{CKMfitter}\\\hline
	$\phi_d$ & $(44.4\pm 1.6)^{\circ}$ & \cite{CKMfitter}\\\hline
	$\lambda$ & $0.225\pm 0.00024$ & \cite{CKMfitter} \\\hline
	$\epsilon$ & $0.05332\pm 0.00011$ & $\epsilon = \lambda^2/(1-\lambda^2)$\\\hline	
\end{tabular} \\
	\caption{\label{tab:data_external}External data input}
\end{table}
	Apart from the ratios of the $BFs$ other observables are treated as statistically independent. Especially, a simple calculation shows that the correlation between asymmetries and branching fractions is small and therefore can be neglected.

\section{Fits to current experimental data}
\label{sect:fits}
To obtain the set of the parameters given in Eqs. \eqref{eq:B2pipi} and \eqref{eq:B2Kpi} a $\chi^2$ minimisation is performed. For each observed ratio of the BFs, Eq. \eqref{eq:chi2_Rcov} is used, while the asymmetries are treated as independent and their $\chi^2$ is given by
\begin{align}
    \chi^2 = \frac{(A_{\text{exp}}-A_{\text{theory}})^2}{\sigma_{\text{exp}}^2}.
\end{align}
The $SU(3)$ limitations, given in Eqs.\eqref{eq:R_TC} and \eqref{eq:SU3_phases} are applied in the same manner.
 
\begin {figure}[t]
                 \centering\includegraphics[width=0.45\textwidth]{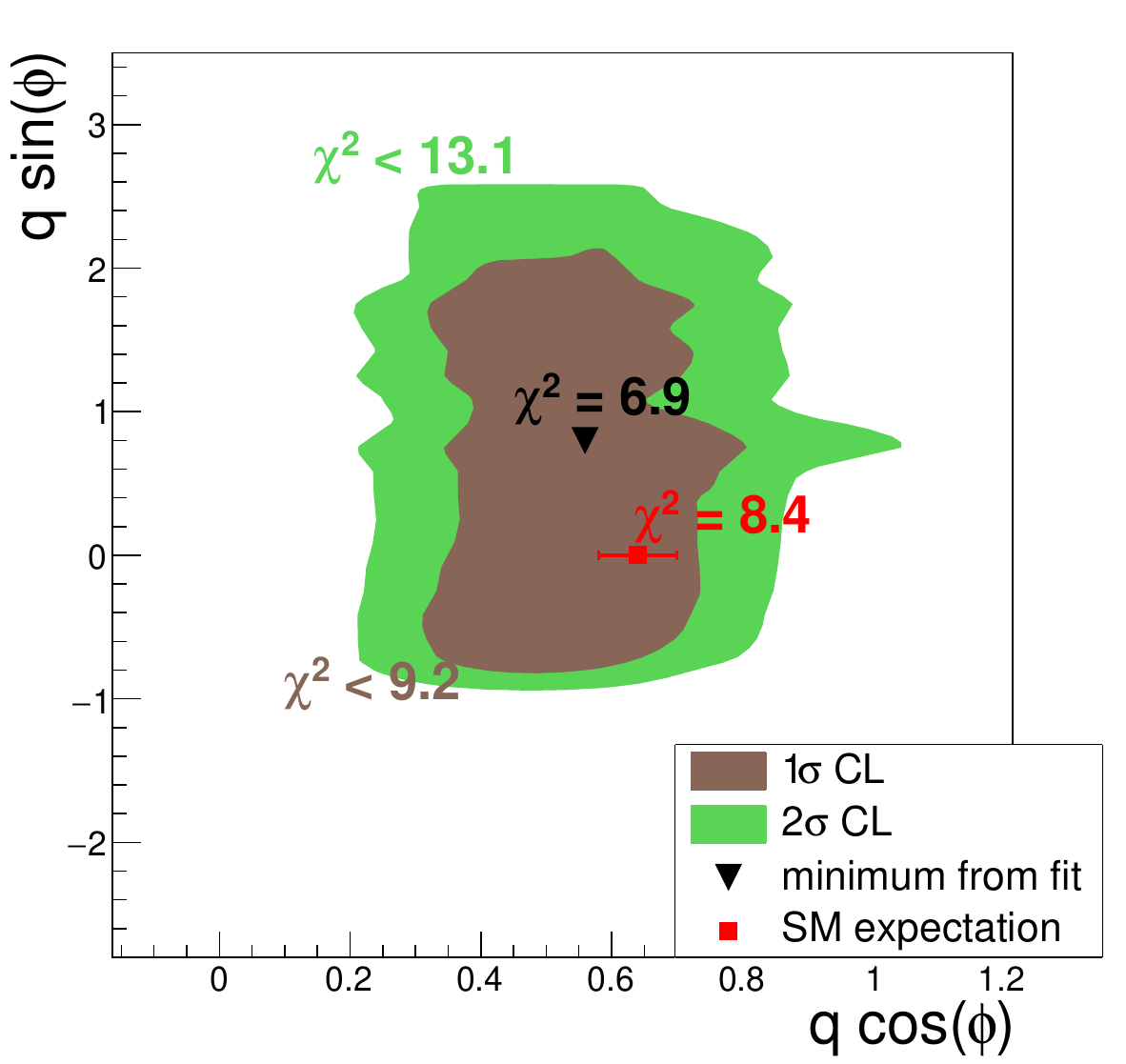}
                 \caption{\label{fig:q_phi_chi2}The $\chi^2$ profile of the $q$-$\phi$ plane with $1\sigma$ and $2\sigma$ contours highlighted. 
		The best-fit point and the $SM$ expectation value are shown.}
	\end{figure}

 The $EW$ penguin amplitudes are expressed by the tree-like input to limit the number of the free parameters. This is done in the frame of $SU(3)$ decomposition (\emph{c.f.} equations~\eqref{eq:EWP_Kpi_SU3}, \eqref{eq:EWPC_SU3} and \eqref{eq:EWPpipi}). The $q-\phi$ parameter plane is scanned obtaining the $\chi^2$ value at each point. The rest of the parameters are profiled. In addition, since $|r_{\rho}|$ is expected to be much smaller than one, a limit on its real part $|\Re{(r_{\rho})}|<0.3$ is imposed. This limit helps in the convergence of the fit and it has been checked that the choice of a value different from $0.3$ has a negligible impact on the values of the extracted parameters.
The results of the scan performed using the Minuit package, \cite{James:310399}, are shown in Fig.~\ref{fig:q_phi_chi2}. The $SM$ expectation lies within the $1\sigma$ region and hence cannot be excluded.   
 
The $1\sigma$ and $2\sigma$ contour values are obtained from the inverse $\chi^2$ distribution, with the use of the minimal value at the best-fit point:
\begin {subequations}
\begin {align}
	&&\text{contour}(1\sigma) &= \chi^2_{min}+(\chi^2)^{-1}(0.68, N_{dof}),\\
	&&\text{contour}(2\sigma) &= \chi^2_{min}+(\chi^2)^{-1}(0.95, N_{dof}).
\end{align}
\end {subequations}
Here, $0.68$ and $0.95$ correspond to $1\sigma$ and $2\sigma$ confidence levels ($CL$) respectively (precise values from a cumulative normal distribution are used to obtain the contours)). 
Number $N_{dof}=2$ is the difference of the number of the degrees of freedom between the nominal fit and the scan that has two parameters fixed.
Apart from the $CL$ contours around the global minimum also the expectation value of the $SM$ is depicted. 

The fact that the $2\sigma$ contour is broad means that there is space for a NP influence at the present experimental status. However, any specific $NP$ prediction concerning $q$ and $\phi$ can be tested using this $\chi^2$ distribution. 
When a model prediction is significantly away from the data minimal point constraints on its parameters may be imposed. For example, on the SMEFT Wilson coefficients~\cite{Grzadkowski:2010es}. By varying one (or more) Wilson coefficient (WC) at the SMEFT scale ($1$~TeV) its reflection on $q$ and $\phi$ can be obtained by matching it onto the WC at the scale of the $B$ meson decays $4.8$~GeV. The $q-\phi$ profile plot (Fig.~\ref{fig:q_phi_chi2}) can be then used to assign $\chi^2$ to the given SMEFT WC value.

The dependence of the result presented in Fig.~\ref{fig:q_phi_chi2} on the observables listed in Section~\ref{sect:data} has been systematically examined.  The strongest sensitivity has been found for the $B^0\to K^0\pi^0$ $BF$ and $CP$ asymmetry. These measurements' precision can be further improved at Belle~II and possibly at LHCb~\cite{LHCb:2020dpr}.

To test the assumption of the $CP$ conserving phase $\omega=0$, the fit is repeated for value $\pm10^{\circ}$ showing no significant change with respect to the nominal fit. 

Furthermore, the sum rule defined in Eq.~\eqref{eq:sumrule} is analysed in the same manner as $q$ and $\phi$. First, the sum rule is made to be a parameter of the fit. Then it is fixed one by one to a range of values. At each point, a fit is performed to the rest of the parameters (including $q$ and $\phi$). Two scenarios are studied. In the $SM$ case $\phi=0$ and $q = 0.64\pm0.06$, while when the $NP$ is allowed these parameters are set free. 
\begin{figure}[b]
    \centering
    \begin{minipage}[b]{0.24\textwidth}
        \centering
        \includegraphics[width=\textwidth]{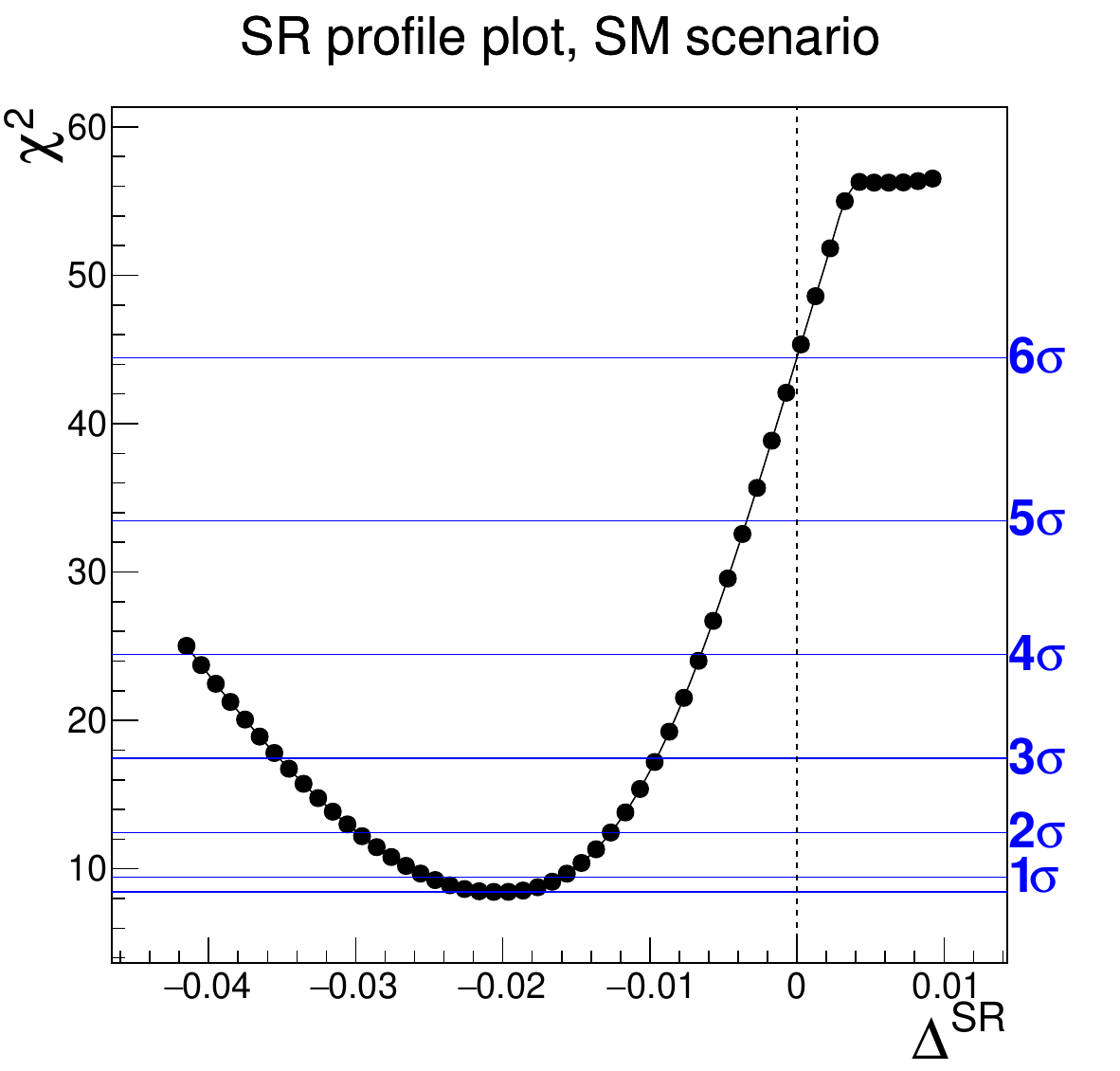}
    \end{minipage}%
    \begin{minipage}[b]{0.24\textwidth}
        \centering
        \includegraphics[width=\textwidth]{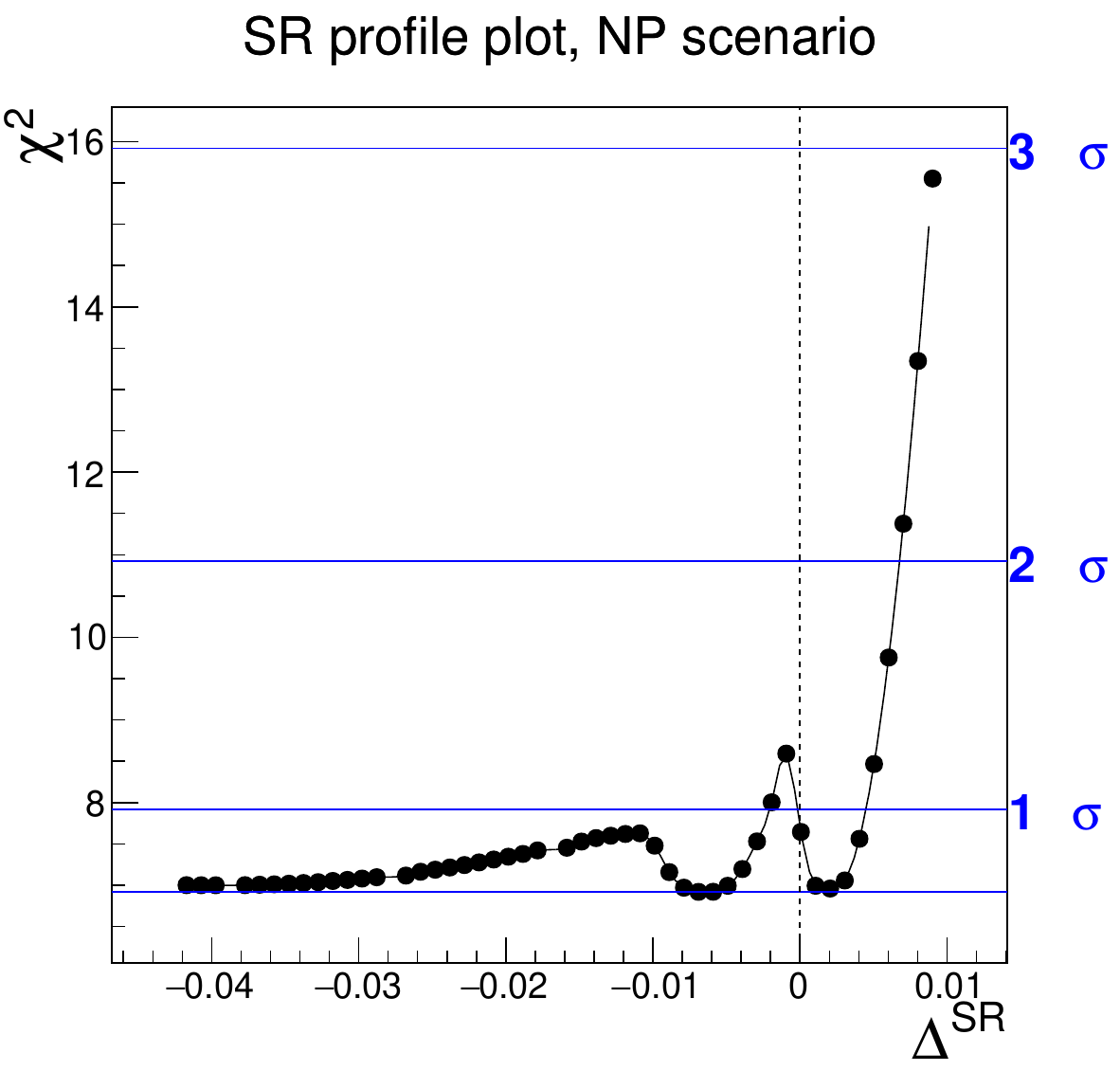} 
    \end{minipage}
    \caption{Profile plots of the sum rule (Eq.~\eqref{eq:sumrule}) for the $SM$ (left) and $NP$ (right) scenario.}
    \label{fig:SR}
    \end{figure}
The results are presented in Fig.~\ref{fig:SR} showing that in the $SM$ scenario, the sum rule value is small (in the scale of an asymmetry), however over $6\sigma$ below zero. This is interesting but not in contradiction with the $SM$ expectations. In the $NP$ case, the sum rule is very weekly constrained from below. From above the $2\sigma$ regions ends at $\Delta^{SR}\lesssim 0.007$, while zero lies within the $1\sigma$ region. The smallness of the obtained values of $\Delta^{SR}$ is not surprising as the applied $SU(3)$ hamiltonian decomposition is used to derive the sum rule, as presented in~\cite{Gronau:1998fn}.
\subsection{The revision of the ``$B\to K \pi$ puzzle''}

The analysis reveals that the precision of the current experimental results is insufficient to constrain the fitted parameters strongly. This corresponds to a rather weak dependence of the $\chi^2$ on these parameters.
The SM prediction lies within the $1\sigma$ confidence region of the $\chi^2$ distribution in the $q-\phi$ plane (see Fig.~\ref{fig:q_phi_chi2}). Therefore, no NP scenarios are required to explain the data. 

The consistency of the SM expectations and the data leads to a conclusion that the ``$B\to K\pi$ puzzle'' is solved. Using the parameterisation of Section~\ref{subsect:par1} and the approximation given in the appendix~\eqref{eq:Kpi_of_parameters} $\Delta A_{CP}$ can be expressed as
\begin{align}
    \Delta A_{CP} \simeq -2\Im(r'_C)\sin\gamma+2\Im (r'_T+r'_C)q\sin\phi.
\end{align}
As seen from Eqs.~\eqref{eq:Kpi_of_parameters}  $A^{\pi^+K^0}$, $A^{\pi^0K^0}$ and $S^{\pi^0K^0}$ are sensitive to $q$ and $\sin{\phi}$, while $BF(\pi^0K^+)$ and $BF(\pi^0K^0)$ depend on $q$ and $\cos{\phi}$. 
 In the SM scenario, the high value of $\Delta A_{CP}$ is assured by large $\Im{(r_C)}$ as presented in Table~\ref{tab:results},
\begin{table}[tbh]
        \centering
        \begin{tabular}{|c|c|c|}
                \hline
                &SM scenario &NP scenario\\\hline
                $\phi$ & $0$& $(55\pm6)^{\circ}$ \\\hline
                $q$ & $0.611\pm0.053$& $0.98\pm 0.10$ \\\hline
                $\gamma$ &$(65.2\pm1.0)^{\circ}$ & $(65.4\pm1.3)^{\circ}$ \\\hline
                $\Im{r_C'}$ & $-0.0566\pm0.0070$ & $-0.079\pm0.011$\\\hline
                $\Im{r_T'}$ & $0.043\pm 0.016$& $-0.022\pm0.011$ \\\hline
                $\boldsymbol{\Delta A_{CP}}$ &$\boldsymbol{0.109\pm0.012}$ & $\boldsymbol{0.107\pm0.013}$\\\hline
                $\boldsymbol{\Delta^{SR}}$ & $\boldsymbol{-0.0226\pm0.0043}$ & $\boldsymbol{-0.0079\pm 0.0045}$\\\hline
        \end{tabular}  
        \caption{The results of the Minuit fits of the main parameters for the SM and NP scenario. Recalculated values of $\Delta A_{CP}$ show agreement with the data.}
        \label{tab:results}
\end{table}
where the uncertainty is obtained by the propagation of errors from the Minuit output.
The results for both scenarios agree with the experimental value given in Eq.~\eqref{eq:DA_CP_exp}. In the SM case the nonzero estimate is driven by the significant strong~phases of the tree amplitudes. At the same time, the obtained uncertainties leave space for the potential NP effects. The sum rule outcomes are presented here for completeness.

The sum rule defined in Eq.~\eqref{eq:sumrule} is approximated by
\begin{align}    
 \Delta^{SR} =& 4q\frac{\Im{(r'_C)}\Re{(r'_T)}-\Im{(r'_T)}\Re{(r'_C)}}{2\Re{(r'_{\rho})}+|r'_{\rho}|^2+1}\sin{(\gamma-\phi)}\\&+\mathcal{O}({r'_T}^3. {r'_C}^3, {r'_{\rho}}^3).
\end{align}
It is strongly limited in the SM scenario, the $2\sigma$ CL amounts to $[-0.0300; -0.0125]$. On the other hand, the sum rule is very weakly constrained when NP is allowed.

\section{Summary}
\label{sect:summary}

Even though the ``$B\to K\pi$ puzzle'' has been solved, the current experimental uncertainties of the analysed decays still leave space for the NP scenarios.
Moreover, the $q-\phi$ profile plot obtained here can serve as a starting point to impose constraints on NP models. 
From the experimental point of view it is clear that improving the precision of the $B^0\to K^0\pi^0$ measurements should help to clarify the picture. 

The sum rule~\eqref{eq:sumrule} result is $6\sigma$ below zero at the minimum of the SM scenario. On the other hand, its value is small in the sense of the asymmetry scale ($\sim2\%$) which is a consequence of the applied parameterisation and the $SU(3)$ decomposition. Adding two more degrees of freedom along with the potential NP weakens the experimental constraints on the sum rule showing compatibility with zero within $1\sigma$. This fact is interesting and calls for a discussion.

As a follow-up, other decays can be examined in a similar analysis. For example $D\to PP'$ or $B\to DP$.

\begin{acknowledgements}
In the first place, I would like to express my gratitude to Professor~Wojciech~Wiślicki who was the initiator of this work and with whom I had many fruitful discussions on the topic. I am deeply grateful to Professor~Stefan~Pokorski, Professor~Janusz~Rosiek, Professor~Mikołaj~Misiak and Dr.~Dibyakrupa~Sahoo for much advice that I received during our consultations. Lastly, special thanks to Dr.~Wojciech~Krzemień for many editorial corrections at the final stage of the writing of this paper.
\end{acknowledgements}
\appendix
\appendix
\section{BF phase-space correction.}
\label{app:corr}
The BF is corrected by the following phase-space factor.
\begin{align}
    BF^{corr}(h_1h_2)&= \frac{M_{B}}{\Phi(\frac{m_{h_1}}{M_{B}},\frac{m_{h_2}}{M_{B}})}\frac{1}{\tau_{B^+}}BF(h_1h_2)\\
    \Phi(X, Y) &= \sqrt{[1-(X+Y)^2][1-(X-Y)^2]}, \nonumber
\end{align}
    Where $h_1, h_2 = \pi^+, \pi^-,\pi^0, K^+, K^-, K^0$\newline and $B = B^+, B^-, B^0$. 

\section{Parameterised predictions for observables}
Within the proposed parameterisation the observables listed Table~\ref{tab:data_Kpi} can be expressed by the following approximations:
\begin{subequations}
\begin{align}
    A^{\pi^+K^-} =&-2\Im{(r'_T)}\sin{\gamma}+2\Im(r'_{\rho})\sin{\gamma}\nonumber\\&+ \mathcal{O}({r'_T}^2, {r'_C}^2, {r'_{\rho}}^2), \\
    A^{\pi^0K^+} =&~2\Im{(r'_{\rho})}\sin{\gamma}-2\Im{(r'_T+r'_C)}(\sin{\gamma}-q\sin{\phi})\nonumber\\&+\mathcal{O}({r'_T}^2, {r'_C}^2, {r'_{\rho}}^2), \\
    A^{\pi^+K^0} =&~2\Im{(r'_{\rho})}\sin{\gamma}-2\Im{(r'_T+r'_C)}(\sin{\gamma}\nonumber \\&-q\sin{\phi})+ \mathcal{O}({r'_T}^2, {r'_C}^2, {r'_{\rho}}^2), \\
    A^{\pi^0K^0} =&~2\Im{(r'_C)}\sin{\gamma} -2\Im{(r'_T+r'_C)}q\sin{\phi}\nonumber \\ &+2\Im{(r_{\rho})}\sin{\gamma}+\mathcal{O}({r'_T}^2, {r'_C}^2, {r'_{\rho}}^2),\\
    S^{\pi^0K^0} =& ~\Big[2\Re{(r'_{\rho})}\cos{\gamma}-1\Big]\sin{\phi_d}\nonumber\\&-\Big[2\Re{(r'_C+r'_{\rho})}\sin{\gamma}+2\Re{(r'_T+r'_C)}q\sin{\phi}\Big]\nonumber\\&\times
    cos{\phi_d}+\mathcal{O}({r'_T}^2, {r'_C}^2, {r'_{\rho}}^2),
    \end{align}
\end{subequations}
\begin{subequations}
    \begin{align}
            BF(\pi^+K^-) =&~2P^2\Big[1-2\Re{(r'_T)}\cos{\gamma\nonumber}\\&\hskip 2cm +\mathcal{O}({r'_T}^2, {r'_C}^2, {r'_{\rho}}^2)\Big],\\
    BF(\pi^+K^0) =&~2P^2\Big[1+2\Re{(r'_{\rho})}\cos{\gamma}\nonumber \\&\hskip 2cm+ \mathcal{O}({r'_T}^2, {r'_C}^2, {r'_{\rho}}^2)\Big],\\
    BF(\pi^0K^+) =&~2P^2\Big[1+2\Re{(r'_\rho-r'_T-r'_C)}\cos{\gamma}\nonumber\\
     &+2\Re{(r'_T+r'_C)}q\cos{\phi}+\mathcal{O}({r'_T}^2, {r'_C}^2, {r'_{\rho}}^2) \Big],\\
    BF(\pi^0K^0) =&~2P^2\Big[1-2\Re{(r'_T+r'_C)}q\cos{\phi}\nonumber\\
    &+2\Re{(r'_C)}\cos{\gamma}+\mathcal{O}({r'_T}^2, {r'_C}^2, {r'_{\rho}}^2)\Big].
    \end{align}\label{eq:Kpi_of_parameters}
\end{subequations}
One finds the expression for the $\Delta A_{CP}$ defined in Eq.~\eqref{eq:DA_CP}:
\begin{align}
    \Delta A_{CP} \simeq -2\Im(r'_C)\sin\gamma+2\Im (r'_T+r'_C)q\sin\phi
\end{align}    

\section {Branching fractions covariance matrix}
\label{app:Cov}
It is clear that some ratios  of BFs are correlated \emph{e.g.} $R^{\pi^0\pi^+}_{\pi^+\pi^-}$ and $R^{\pi^0\pi^+}_{\pi^0K^+}$. To account for this correlation, a covariance matrix is formulated with the use of the following approximated equalities:
	\begin {subequations}
	\begin {align}
		\left\langle \left(\frac{Y}{Z}\right)^2\right\rangle - \left\langle \left(\frac{Y}{Z}\right)\right\rangle^2 =~& \sigma_Y^2\left(\frac{1}{Z^2}+\frac{3\sigma_Z^2}{Z^4}\right)\\&+Y^2\left(\frac{\sigma_Z^2}{Z^4}+\frac{\sigma_Z^4}{Z^6}\right),\\
		\left\langle \frac{X}{Z}\frac{Y}{Z}\right\rangle-\left\langle\frac{X}{Z}\right\rangle\left\langle\frac{Y}{Z}\right\rangle =~& XY\frac{\sigma_Z^2}{Z^4}\left(1-\frac{\sigma_Z^2}{Z^2}\right),\\
		\left\langle\frac{Z}{X}\frac{Y}{Z}\right\rangle-\left\langle\frac{Z}{X}\right\rangle\left\langle\frac{Y}{Z}\right\rangle  =~& -\frac{X}{Y}\frac{\sigma_Z^2}{Z^2}\left(1+\frac{\sigma_Y^2}{Y^2}\right),\\
		\left\langle\frac{Z}{X}\frac{Z}{Y}\right\rangle -\left\langle\frac{Z}{X}\right\rangle\left\langle\frac{Z}{Y}\right\rangle  =~& \frac{\sigma_Z^2}{XY}\Bigg(1+\frac{\sigma_X^2}{X^2}\nonumber\\&+\frac{\sigma_Y^2}{Y^2}+\frac{\sigma_X^2}{X^2}\frac{\sigma_Y^2}{Y^2}\Bigg).
	\end{align}
	\end{subequations}
\bibliographystyle{spphys} 
\bibliography{bibliography}

\end{document}